\documentclass[iop]{emulateapj}
\usepackage{amsmath}
\usepackage{relsize}
\usepackage{hyperref}
\usepackage{color}
\usepackage{listings}
\usepackage{multirow}
\usepackage{sidecap}
\usepackage{hyperref}
\usepackage{float}
\usepackage{xspace}
\usepackage{gensymb}

\begin{document}

\newcommand{\numLRG}{16,191,145\xspace}
\newcommand{\stellarContamination}{1.8\xspace}

\newcommand{\density}{1,426\xspace}
\newcommand{\nside}{512}
\newcommand{\resolution}{$6.9'$}
\newcommand{\pixArea}{$1.31\times10^{-2}\textrm{\,deg}^2$\xspace}

\title{Photometric Selection of a Massive Galaxy Catalog with $z\geq0.55$}

\author{Carolina N\'u\~nez}
\affil{McWilliams Center for Cosmology, Department of Physics, Carnegie Mellon University, Pittsburgh, PA 15213, USA}
\affil{Department of Astrophysical Sciences, Peyton Hall, Princeton University, Princeton, NJ 08544, USA}
\affil{Instituto de Astrof\'isica, Pontificia Universidad Cat\'olica de Chile, Av. Vicu\~na Mackenna 4860, 782-0436 Macul, Santiago, Chile}
\email{cnunez@andrew.cmu.edu}

\author{David N. Spergel}
\affil{Department of Astrophysical Sciences, Peyton Hall, Princeton University, Princeton, NJ 08544, USA}

\author{Shirley Ho}
\affil{McWilliams Center for Cosmology, Department of Physics, Carnegie Mellon University, Pittsburgh, PA 15213, USA}

\begin{abstract}
\hspace{0.5cm}
We present the development of a photometrically selected massive galaxy catalog, targeting Luminous Red Galaxies (LRGs) and massive blue galaxies at redshifts $z\geq 0.55$.  Massive galaxy candidates are selected using infrared/optical color-color cuts, with optical data from Sloan Digital Sky Survey (SDSS) and infrared data from ``unWISE'' forced photometry derived from the Wide-field Infrared Survey Explorer (WISE).  The selection method is based on previously developed techniques to select LRGs with $z>0.5$, and is optimized using receiver operating characteristic (ROC) curves.  The catalog contains \numLRG objects, selected over the full SDSS DR10 footprint.  The redshift distribution of the resulting catalog is estimated using spectroscopic redshifts from the DEEP2 Galaxy Redshift Survey and photometric redshifts from COSMOS.  Restframe $U-B$ colors from DEEP2 are used to estimate LRG selection efficiency.  Using DEEP2, the resulting catalog has average redshift $z=0.65$, with standard deviation $\sigma = 2.0$, and average restframe $U-B=1.0$, with standard deviation $\sigma=0.27$.  Using COSMOS, the resulting catalog has average redshift $z=0.60$, with standard deviation $\sigma = 1.8$.  We estimate $34\%$ of the catalog to be blue galaxies with $z\geq0.55$.  An estimated $9.6\%$ of selected objects are blue sources with redshift $z<0.55$.  Stellar contamination is estimated to be $\stellarContamination\%$.

\end{abstract}


\section{Introduction}
\label{sec:Introduction}

Luminous Red Galaxies (LRGs) are particularly suited to the study of clusters.  These elliptical galaxies are typically the most luminous and massive galaxies at redshifts $z\le1.0$, strongly tracing their underlying dark matter halos.  Furthermore, their uniform spectral energy distribution (SED) and characteristic spectral features have allowed for simplified selection and accurate redshift determination at $z<0.5$ ~\citep{Eisenstein:2001aa, Padmanabhan:2005aa}.  This previous work takes advantage of a strong break in the SED of LRGs that occurs at $4000\,\textrm{\AA}$.  As objects at higher redshift are considered however, this method becomes limited, as the $4000\,\textrm{\AA}$ feature passes into the \textit{i} band at $z\sim0.75$.  In order to efficiently select LRGs at higher redshifts, new techniques must be used.

This paper presents a publicly available catalog of massive galaxy candidates, with redshifts  $z\geq0.55$.  Galaxies are chosen based on photometric selection methods aimed to select higher redshift LRGs by combining optical and infrared surveys, developed by ~\citet{Schlegel:2011aa} and further improved by~\citet{Prakash:2015aa}.   This work extends these previous results to include massive blue galaxies in addition to LRGs, and optimizes the selection cut using receiver operating characteristic (ROC) curves to target objects above $z\geq0.55$.  Optical data used is from the Sloan Digital Sky Survey~\citep[SDSS;][]{York:2000aa} and infrared data used is from a catalog of forced photometry derived from the Wide-field Infrared Survey Explorer~\citep[WISE;][]{Wright:2010aa,Lang:2014aa}.  The resulting catalog's redshift distribution is tested using spectroscopic and photometric redshifts, and restframe $U-B$ are used to determine LRG selection efficiency.  A full description of the data sets used in this work is presented in \S\ref{sec:Data}.  Selection technique and optimization are discussed in \S\ref{sec:Selection}.  The properties of the resulting catalog, and its comparison to similar work by~\citet{Prakash:2015aa}, are described in \S\ref{sec:Results}.

\section{Data}
\label{sec:Data}

\subsection{SDSS/WISE Forced Photometry}
\label{subsec:unWISE}

Infrared and optical data used in this work are provided by the publicly available\footnote{\href{http://unwise.me}{http://unwise.me}} forced photometry catalog developed by~\citet{Lang:2014aa}.  The catalog provides improved photometry for WISE, at the positions of over 400 million primary sources from SDSS-III Data Release 10~\citep{Eisenstein:2011aa, Ahn:2014tenth}.  Taking advantage of the higher resolution of SDSS as a means to find sources in WISE, photometry is extracted using \textit{The Tractor} image modeling code from ``unWISE'' unblurred coadds of WISE imaging.  This unWISE imaging preserves the original resolution of the survey, and allows for a higher signal-to-noise.  For a detailed overview of the ``unWISE'' imaging, see~\citet{Lang:2014ab}.  The resulting catalog provides a consistent set of sources between SDSS and WISE.  Forced photometry may also be obtained for sources which, although blended in WISE, are resolved in SDSS.

The SDSS was conducted using a dedicated 2.5-meter telescope at the Apache Point Observatory, New Mexico.  The telescope is equipped with two multi-object spectrographs as well as a 120-megapixel wide-field camera~\citep{Gunn:1998aa} performing 5-band \textit{ugriz} photometry at wavelengths 3551, 4686, 6165, 7481, 8931 $\AA$~\citep{Gunn:2006aa,Fukugita:1996aa}.  To date, SDSS has made public twelve data releases \citep{abazajian2003first, abazajian2004second, abazajian2005third, adelman2006fourth, adelman2007fifth, adelman2008sixth, abazajian2009seventh, aihara2011eighth, ahn2012ninth, Ahn:2014tenth,Alam:2015eleventhtwelfth}.  This work uses the $r$ and $i$ bands.

WISE is a full-sky cryogenic survey, carried out in 2010 over four simultaneously observing bands centered at 3.4, 4.6, 12, and 22 $\mu\textrm{m}$.  The survey achieved unprecedented sensitivity and angular resolution over the full sky; the most recent AllWISE data release attained 5$\sigma$ point source sensitivities better than 0.054, 0.071, 0.730 and 5 mJy in unconfused regions on the ecliptic, where sources can be distinguished from background noise, as well as 6.1", 6.4", 6.5", and 12.0" angular resolution, in each of the four bands.  For a full description of WISE, see~\citet{Wright:2010aa}.  This work uses the $3.4\,\mu\textrm{m}$ (W1) band.

We use objects in the unWISE catalog that are marked as galaxies in the SDSS \textit{PhotoObj} catalog files, with clean photometry flag set to 1\footnote{\href{https://www.sdss3.org/dr10/tutorials/flags.php\#cleanflag}{https://www.sdss3.org/dr10/tutorials/flags.php\#cleanflag}}.  The catalog is then masked to ensure we are only using regions with good SDSS observing conditions: objects not contained within imaging masks provided by~\citet{ShirleyMask} are excluded from the analysis.  The photometry is then corrected for galactic extinction following~\citet{Schlegel:1998aa}.  Values of $E(B-V)$ are obtained from dust maps provided by~\citeauthor{Schlegel:1998aa}, and extinction value $A(\lambda)$ for each band is calculated using $A/E(B-V)$ of 2.751, 2.086, and 0.234 for $r$, $i$, and W1, respectively\footnote{\href{https://github.com/astropy/astroquery/blob/master/astroquery/irsa_dust/tests/data/dust_ext_detail.tbl}{https://github.com/astropy/astroquery/blob/master/\\astroquery/irsa\_dust/tests/data/dust\_ext\_detail.tbl}}.  Extinction corrected magnitudes are then given by the apparent magnitude, subtracted by the extinction value.

In the optical, $r$ and $i$ band magnitudes are assumed to be consistent with AB magnitudes.  In the infrared band, WISE apparent magnitudes are converted from Vega to AB magnitudes, using the following magnitude offset in the W1 band\footnote{\href{http://wise2.ipac.caltech.edu/docs/release/allsky/expsup/sec4_4h.html\#conv2ab}{http://wise2.ipac.caltech.edu/docs/release/allsky/expsup/ \\ sec4\_4h.html\#conv2ab}}.

\begin{equation*}
\label{m_AB}
m_{AB} = m_{Vega} + 2.699
\end{equation*}

We limit the $r$, $i$, and W1 band (AB) magnitudes to 22.9, 21.8, and 20.5, respectively.  Magnitude errors in each band are restricted to values below 0.2, corresponding to a 5\,$\sigma$\ detection.  In regions of the surveys with shallower photometry, it is more difficult to reach this limit; these cuts perform best in regions with deeper photometry (e.g. Stripe 82).

\subsection{DEEP2 Galaxy Redshift Survey}
\label{subsec:DEEP2}

The DEEP2 Galaxy Redshift Survey is the second phase of the Deep Extragalactic Evolutionary Probe (DEEP) survey conducted using the Keck II telescope, utilizing the DEIMOS spectrograph~\citep{Faber:2003aa}.  The most recent fourth data release contains all data from previous releases, providing spectra for $\sim$50,000 galaxies, selected using BRI optical catalogs by \citet{Coil:2004aa}.  The DEEP2 DR4 Redshift Catalog contains 50,319 unique entries with $0.0 < z < 1.4$, and covering $2.8\,\textrm{deg}^2$ over four separate 120' by 30' fields.  In particular, we use the field centered at $14^{\textrm{h}}17^{\textrm{m}}$, +$52\degree30'$, coincident with the Extended Groth Strip (EGS).  The redshift catalog also contains $U-B$ restframe colors obtained from \citet{Willmer:2006aa}.   We consider redshift value \textit{ZBEST}, which is corrected for heliocentric motion, and restframe $U-B$ color, for sources with \textit{ZQUALITY} flag of 3 or 4 that are cross-matched to within 10'' of objects in the SDSS/WISE catalog.  For more details on the DEEP2 survey, see~\citet{Newman:2012aa,Davis:2003aa,Davis:2007aa}.

Furthermore, we make use of a 2D selection function map of the EGS region to determine the completeness of the estimated properties of our sample.  The map contains the probability that an object meeting the DEEP2 target-selection criteria is selected for observation and successfully yielded a redshift.  The selection function maps are described in \citet{Cooper:2006aa}, \citet{Coil:2008aa}, and \citet{Newman:2012aa}. 

\subsection{COSMOS}
\label{subsec:COSMOS}

\subsubsection{Photometric Redshift Catalog}
The Cosmological Evolution Survey (COSMOS) Photometric Redshift Catalog is an accurate, magnitude-limited photo-z redshift catalog extending to $I < 25$.  Redshifts are calculated using 30 bands in the UV (Galaxy Evolution Explorer), visible near-IR (Subaru, CFHT, United Kingdom Infrared Telescope, and National Optical Astronomy Observatory), and mid-IR (Spitzer/IRAC), over a $2\,\textrm{deg}^2$ region of the sky.  For more information, see~\citet{COSMOS1}.

We consider redshift value \textit{zp\_best} for sources that are cross-matched to within 10'' of objects in the SDSS/WISE catalog. To ensure reliable redshifts, we consider only objects in the photo-z catalog where $flagB=0$, $flagV=0$, $flagi=0$, $flagz=0$, and $flagD=0$ to be sure that the objects are not masked in any optical band.  Furthermore, we limit the photo-z catalog to $i^+<24$.

\subsubsection{Zurich Structure and Morphology Catalog}
\label{subsubsec:Zurich}

We use star-galaxy classifications developed in the 2006 May release of~\citet{ACS}, contained within the COSMOS Zurich Structure and Morphology catalog and described in~\citet{Zurich2, Zurich1}.  We select a subset of the catalog where $ACS\_CLEAN = 1$, and cross-match sources to within 10'' of objects in the SDSS/WISE catalog.  The $ACS\_MU\_CLASS$ flag is used as an indicator of star-galaxy classification.

\section{Selection}
\label{sec:Selection}

\begin{figure}
\begin{center}
 \includegraphics[width=\columnwidth]{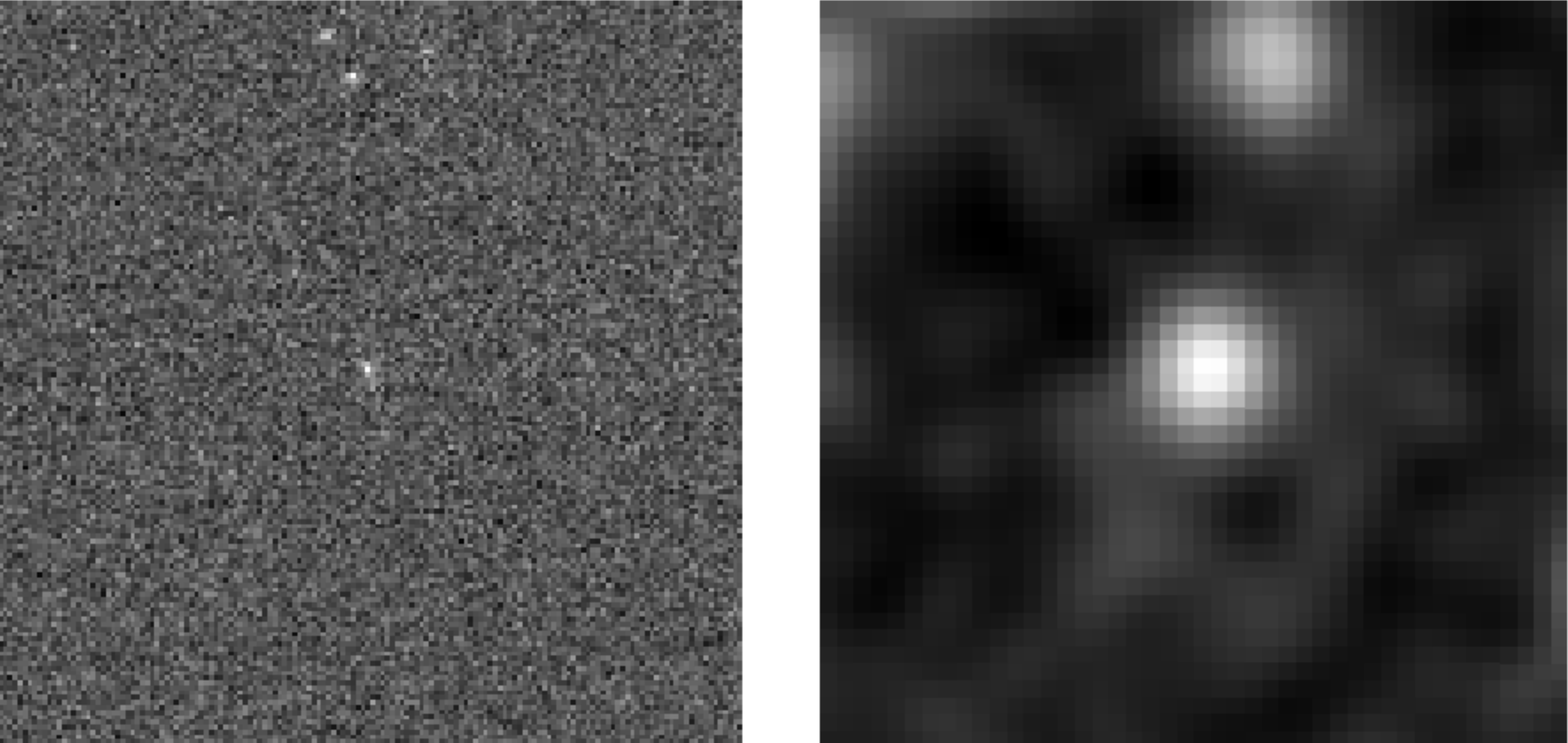}

\caption{As in Figure 3 of \citet{Prakash:2015aa}: An example LRG at $z\approx1$ in SDSS and WISE.
The ``$1.6\,\mu\textrm{m}$ bump'' manifests itself as a peak in infrared-to-optical flux for LRGs at $z\approx1.0$.
Centered in each arcmin squared image, the LRG is shown as seen in SDSS DR10 image Single Field Search\footnote{\href{http://dr10.sdss3.org/fields}{http://dr10.sdss3.org/fields}} in the $r$ band (left), and WISE Image Service\footnote{\href{http://irsa.ipac.caltech.edu/applications/wise/}{http://irsa.ipac.caltech.edu/applications/wise/}} in the $3.4\,\mu\textrm{m}$ W1 band (right).
This property allows for selection of LRGs with $z>0.5$ through a simple cut in $r-3.4\,\mu\textrm{m}$ vs $r-i$.}
 
\label{r_and_w_images}
\end{center}\end{figure}

\subsection{Method}
\label{subsec:Method}

\begin{figure}
\begin{center}
\includegraphics[width=\columnwidth]{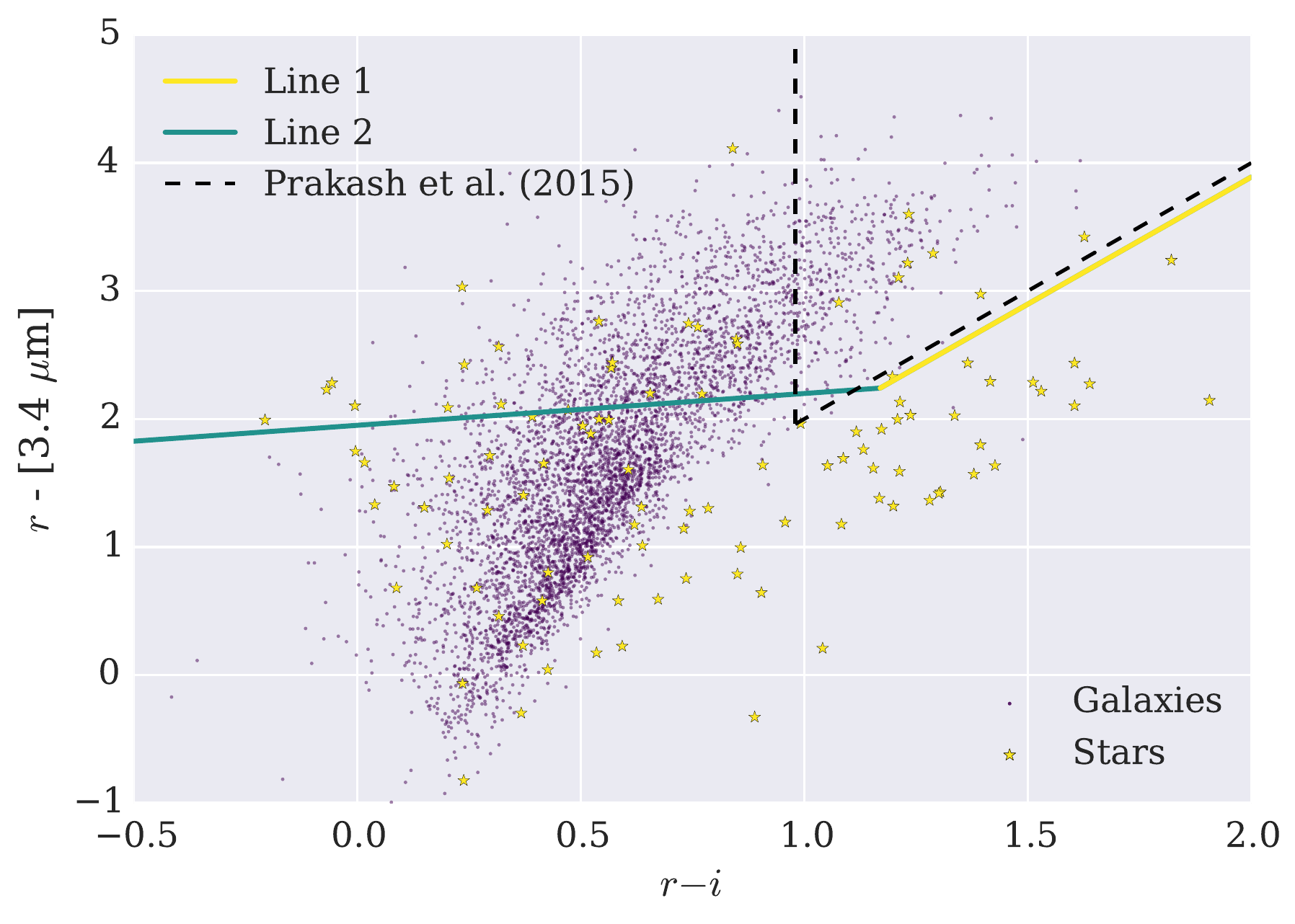}
\caption{Star/Galaxy separation in optical/infrared color-color space.  We optimize a color-color cut targeting objects with $z\ge0.55$, comprised of two lines (Line 1 and Line 2) that are varied as discussed in \S\ref{sec:Selection}.
The figure shows 5039 objects that are cross-matched with the COSMOS Zurich Structure and Morphology catalog, colored by star/galaxy classification, of which 1648 are selected (above Line 1 and Line 2) by the cut.
Photometry is provided by the ``unWISE'' forced photometry catalog, with (AB) magnitudes $r<22.9$, $i<21.8$, $3.4\,\mu\textrm{m}<20.5$ and their respective errors below 0.2.
Stellar contamination is found to be less than $2\%$.
For comparison, a similar selection cut targeting $z>0.6$, by~\citet{Prakash:2015aa} is shown.}
\label{split_by_stargal}
\end{center}\end{figure}

\begin{figure}
\begin{center}
\includegraphics[width=\columnwidth]{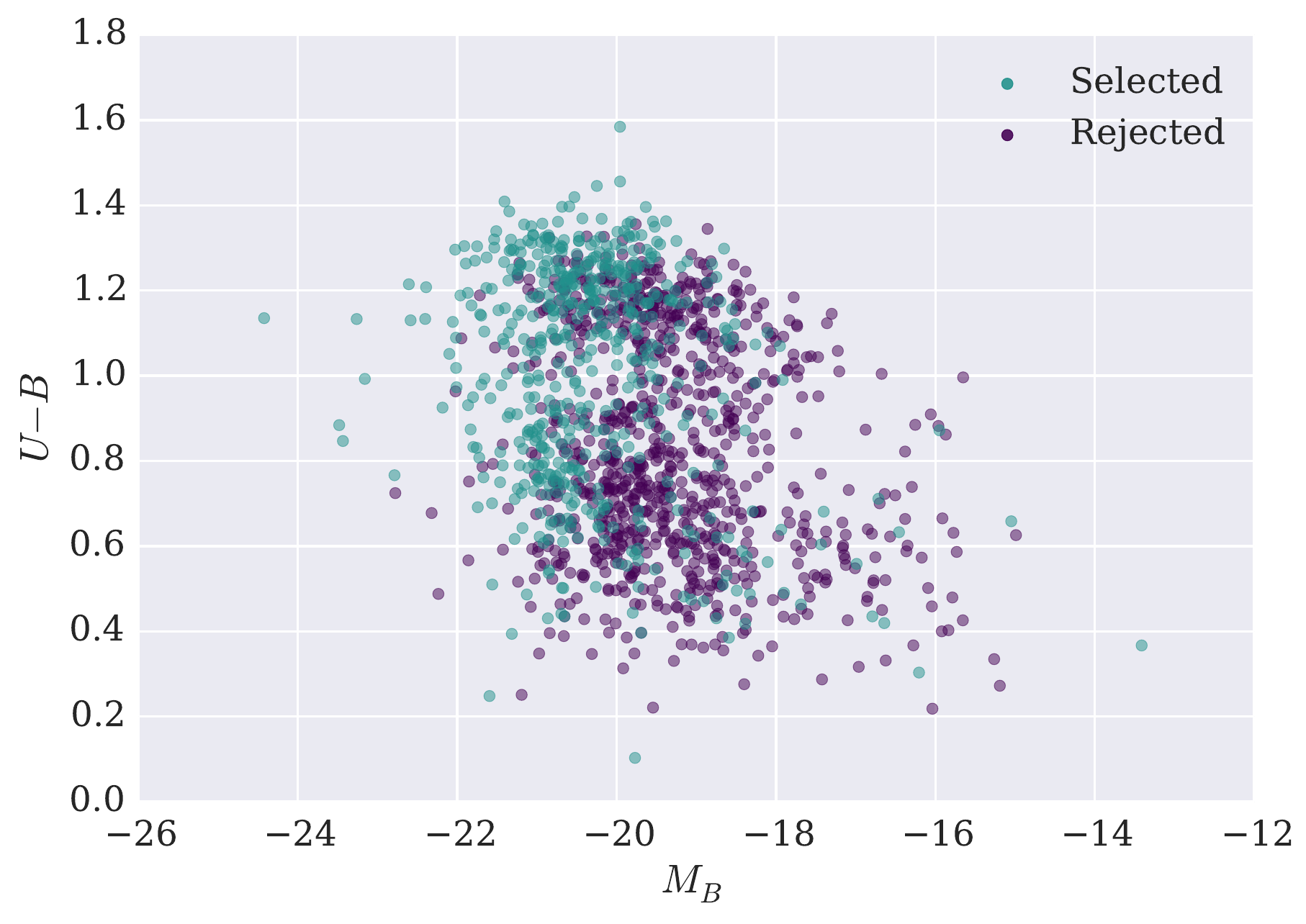}
\caption{Color-magnitude diagram.
The figure shows restframe $U-B$ vs $M_B$ absolute $B$ band (AB) magnitudes provided by the DEEP2 Redshift Survey catalog along the EGS field, for objects contained in SDSS and WISE.
Two populations are visible, corresponding to red and blue sequence galaxies, with red galaxies lying above $U-B > 1.0$.
Objects selected by the cut correspond to more massive objects with red and blue color.}
\label{color-mag-plot}
\end{center}\end{figure}

We seek to photometrically select a catalog of massive galaxies with redshift $z\ge0.55$, building on the LRG selection technique described by~\citet{Schlegel:2011aa} and~\citet{Prakash:2015aa}.  Unable to rely on the $4000\,\textrm{\AA}$ spectral feature used in previous LRG selection techniques~\citep[e.g.,][]{Eisenstein:2001aa}, this new method takes advantage of the presence of cool, old stellar populations and low star formation of LRGs at $0.5<z<1$.  At these redshifts, LRGs exhibit a spectral feature known as the ``$1.6\,\mu\textrm{m}$ bump''.  At restframe wavelength $\lambda_0=1.6\,\mu\textrm{m}$, the SED of cool, old stars exhibit a local maximum due to a local minimum opactity of $\textrm{H}^{-\parallel}$ ions~\citep{John:1988aa}; this is observable, for LRGs at redshifts $z\sim 0.5$--$1$, as a peak in infrared to optical flux at wavelengths of \mbox{$\sim$\,2--4$\,\mu\textrm{m}$}~\citep{Sawicki:2002aa}.  Figure~\ref{r_and_w_images} shows an example LRG as observed in the $r$ and W1 bands.  By combining optical and infrared imaging data, color-color cuts can therefore be made to select high redshift LRGs.

Analysis by~\citet{Schlegel:2011aa} tests this selection technique using the All-Wavelength Extended Groth Strip (EGS) International Survey~\citep[AEGIS;][]{Davis:2007aa}, which provides deep imaging data over all wavelengths in the EGS.  The data set includes publicly available optical imaging from the Canada-France Hawaii Telescope Legacy Survey~\citep[CFHT LS;][]{Gwyn:2008aa}, infrared imaging from the \textit{Spitzer} Infrared Array Camera~~\citep[IRAC;][]{Barmby:2008aa} and redshifts and restframe $U-B$ colors from the DEEP2 Redshift Survey~\citep{Davis:2003aa} to test the selection technique.

The adopted method outlined in~\citeauthor{Schlegel:2011aa} can be summarized as a simple cut in $r-[3.6\,\mu\textrm{m}]$ vs $r-i$, both selecting LRGs at the desired redshift and eliminating galaxies with bluer SEDs.  Here, \textit{r} and \textit{i} represent the optical \textit{r} and \textit{i} bands of the CFHT LS.  The selection proposes $r$ and $i$ band cutoffs of 22.5 and 21.5, respectively, and yields 420 objects per square degree with $[3.6\,\mu\textrm{m}]< 18.9$ and 1120 objects per square degree with $[3.6\,\mu\textrm{m}]< 19.4$, based on a $0.4\textrm{\,deg}^2$ area within the EGS, with 10-15\% uncertainty due to cosmic variance.

As described by~\citet{Schlegel:2011aa}, the lowest band in WISE at $3.4\,\mu\textrm{m}$ is particularly suited to this color-color cut, as it coincides with the $1.6\,\mu\textrm{m}$ bump at $z\sim1$.  Furthermore, a cut in this $r-[3.4\,\mu\textrm{m}]$ vs $r-i$ color-color space also allows for the separation of stars and galaxies, in order to select of a catalog with low stellar contamination.

\citet{Prakash:2015aa} presents a thorough analysis of these methods using optical photometry from CFHT LS, and infrared photometry from WISE.  The cut was optimized to select LRGs with $z>0.6$, by varying the intersection of a vertical line with a sloped line  using the ROC curve and Figure of Merit (FOM) statistics.  The result of this analysis can be summarized by two requirements:
\begin{equation*}
  \begin{gathered}
	r-i>0.98 \\
	r-[3.4\,\mu\textrm{m}] > 2.0 \times (r-i)
\end{gathered}  
\end{equation*}

The analysis in \citet{Prakash:2015aa} is intended to be used for target selection of LRGs in spectroscopic surveys such as the Extended Baryon Oscillation Spectrscopic Survey (eBOSS) and the Dark Energy Spectroscopic Instrubent (DESI) survey.  For further details on eBOSS target selection of LRGs using these cuts, see~\citet{Prakash:2016aa}.

The work presented in this paper uses extinction corrected infrared data from the $3.4\,\mu\textrm{m}$ band of WISE forced photometry, and extinction corrected $r$ and $i$ bands from SDSS.  Model magnitudes are used to calculate $r-i$.  In the case of $r-[3.4\,\mu\textrm{m}]$, however, the $treated\_as\_pointsource$ flag from~\citet{Lang:2014aa} indicates which $r$ band magnitude to use.  If WISE objects were treated as point sources for the purposes of forced photometry, we use PSF magnitudes in the $r$ band.  For those which are treated as extended objects, we use composite model (cmodel) magnitudes in the $r$ band.

\begin{figure}
\centering
\begin{tabular}{l}
    \centering\includegraphics[width=\linewidth]{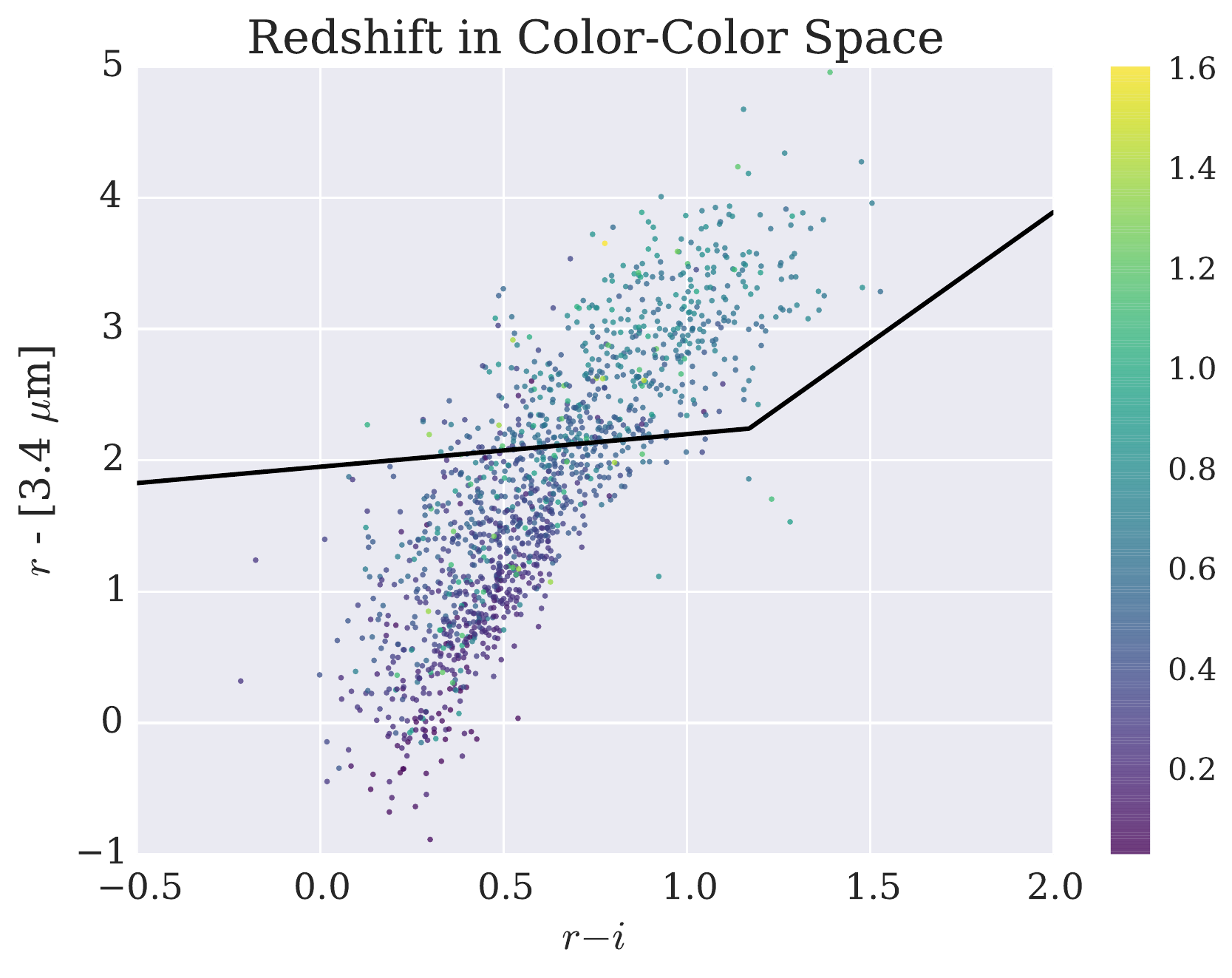}
  \end{tabular}
  \quad
    \begin{tabular}{l}
    \centering\includegraphics[width=\linewidth]{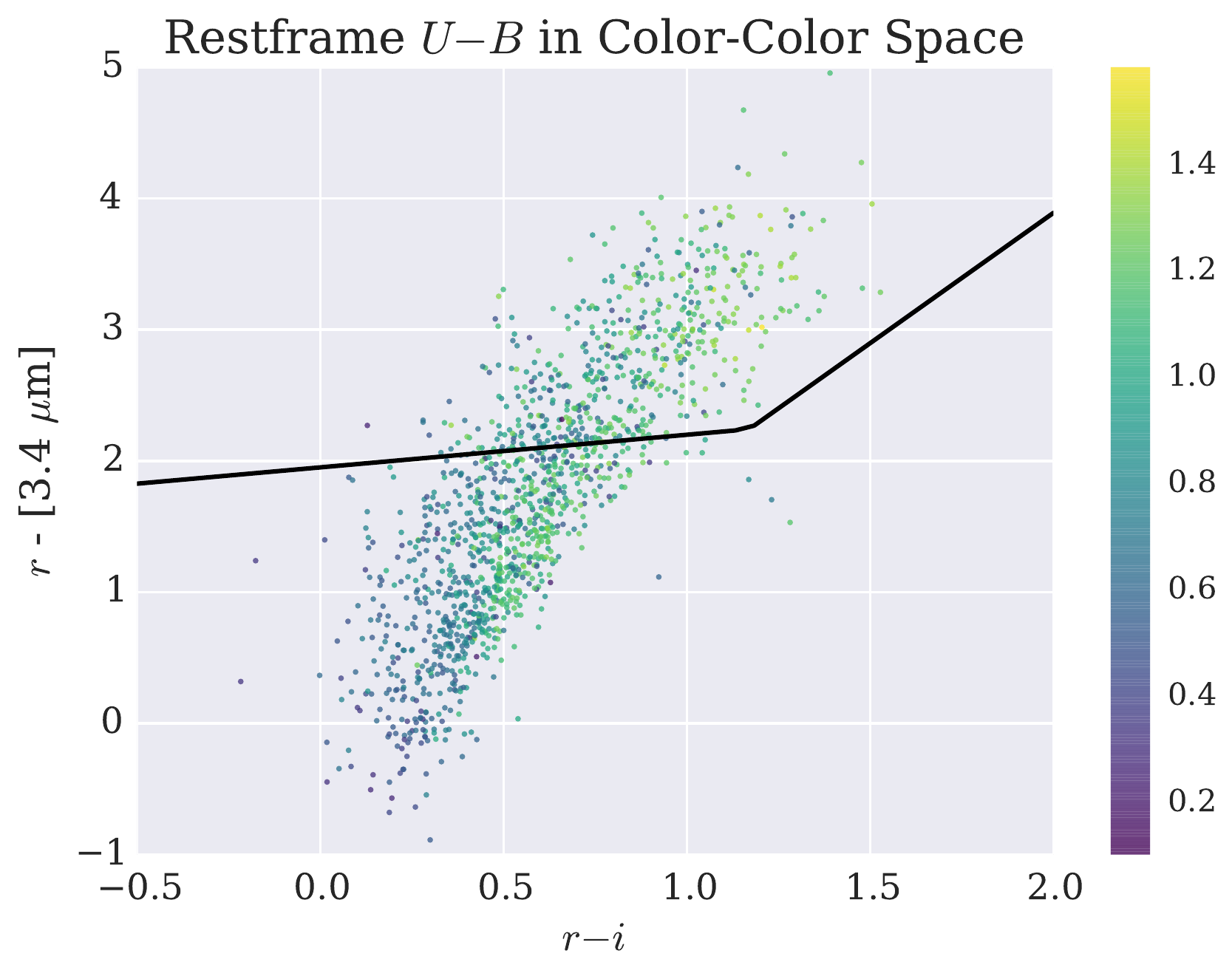} 
  \end{tabular}
  \caption{Properties along optical/infrared color-color space.
Both figures show 1407 objects observed by SDSS, WISE, and the DEEP2 EGS field.
Note that stellar contamination does not appear, as DEEP2 avoided targeting stars.
Top: Objects are colored by spectroscopically measured redshift.
Objects with $z\ge0.55$ are targeted by the selection cut.
Bottom: Objects are colored by restframe $U-B$.  Galaxies with $U-B>1.0$ have LRG-like SED.}
  \label{fig:split}
\end{figure}

In Figure~\ref{split_by_stargal}, the final selection cut is shown, as well as star-galaxy separation.  The color-magnitude diagram of Figure~\ref{color-mag-plot} shows the separation of red and blue sequence galaxies with restframe $U-B$.  In Figure~\ref{fig:split}, we show how the redshift and restframe $U-B$ vary across the color-color space, where restframe $U-B > 1.0$ are indicative LRG-like SED.

\subsection{Optimization}
\label{subsec:Optimization}

 \begin{figure}
  \centering
  \begin{tabular}{l}
    \centering\includegraphics[width=\linewidth]{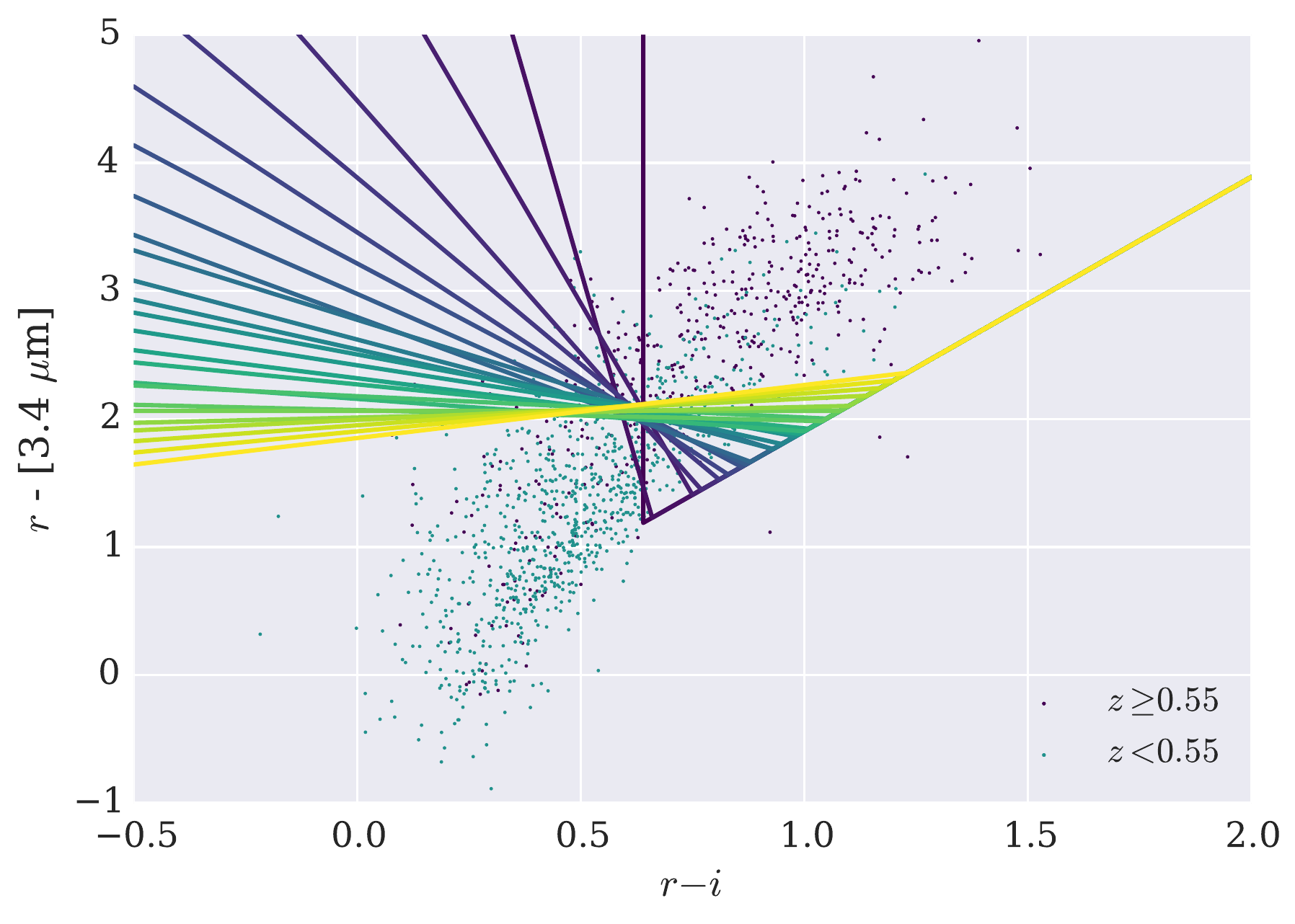}
  \end{tabular}
  \quad
    \begin{tabular}{l}
    \centering\includegraphics[width=\linewidth]{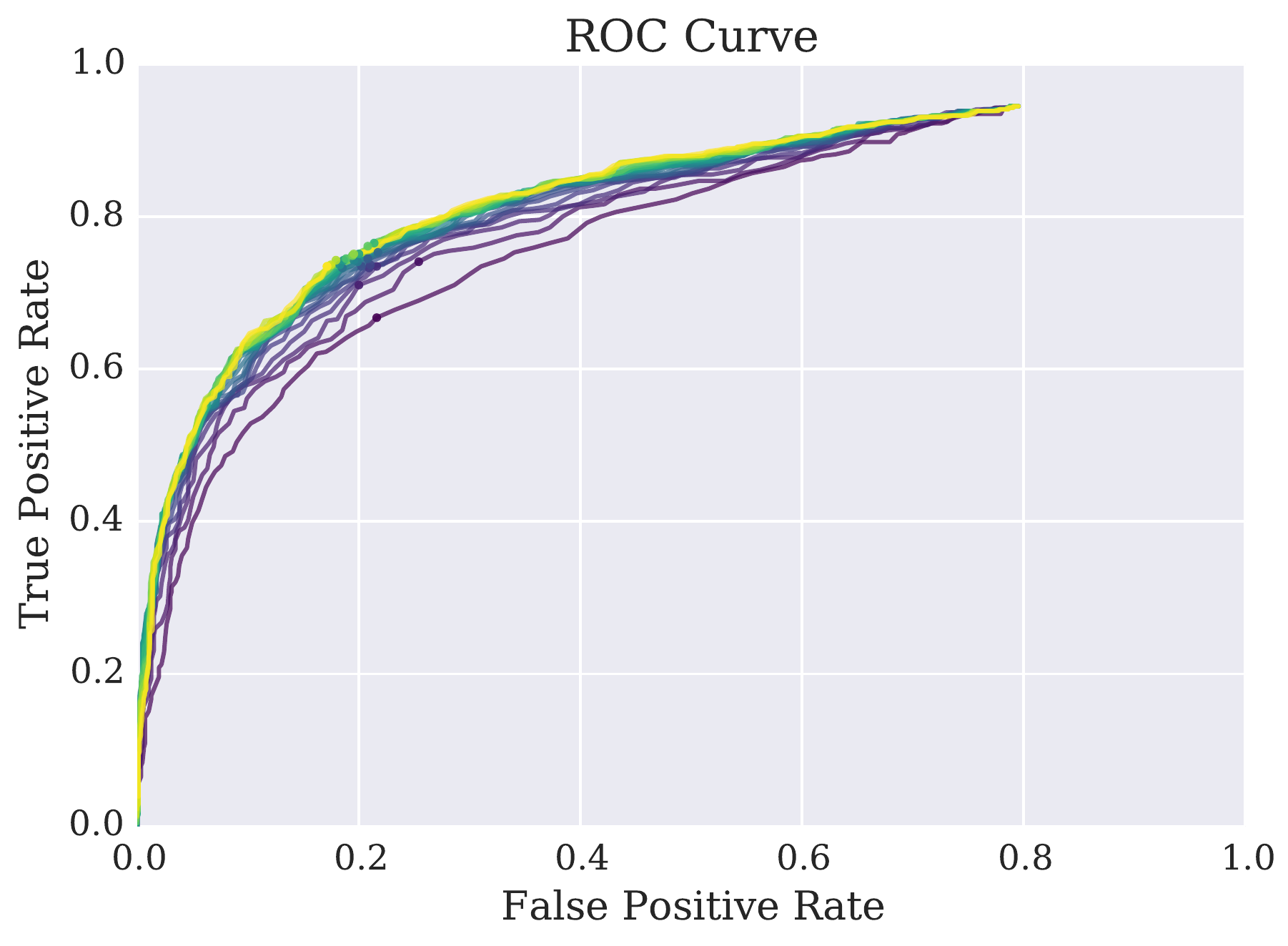} 
  \end{tabular}
\caption{Example classifiers (top) and their corresponding ROC curves (bottom).
As described in \S\ref{subsec:Optimization}, we optimize a color-color cut to select galaxies with $z\ge0.55$.
Top:
We show the optimal line for each slope tested.
For a given slope of the left-hand side line, we generate a ROC curve by calculating the True Positive Rate and False Positive Rate as the point of intersection with the right-hand side line is varied from $r-i=-0.5$ to $r-i=2.0$.
Bottom:
A ROC curve is shown for each slope tested.
The optimal point of intersection for a given slope is given by the point whose distance from (0,1) in the ROC space is minimized.
The best performing classifier is the one whose area under the curve is greatest.}
  \label{fig:ROC}
\end{figure}

We optimize the selection method to select a sample with redshift $z\geq0.55$, and to extend the cut to allow for massive blue galaxies, while maintaining low stellar contamination.  To identify the optimal color-color cut, we use ROC curves (True Positive Rate vs False Positive Rate).  This curve provides a useful statistic to measure the performance of a binary classifier as a threshold is varied. Here, the classifier is a simple cut in $r-[3.4\,\mu\textrm{m}]$ vs $r-i$ yielding two outcomes: above the cut, objects are selected, and below the cut objects are rejected.

A ``true positive'' is defined as an object that has been correctly selected by the classifier, whereas a ``false positive'' is an object that has been incorrectly selected by the classifier.  In this case, an object that is selected by the color-color cut and has an observed redshift $z\geq0.55$ is a true positive; an object that is selected by the color-color cut and has an observed redshift $z<0.55$ is a false positive.  The ``true positive rate'' (also ``completeness'') is the proportion of true positives to the total number of objects being classified that have redshift $z\geq0.55$.  The ``false positive rate'' is the proportion of false positives to the total number of objects being classified that have redshift $z<0.55$. We use 1360 objects from the unWISE catalog cross-matched to DEEP2 spectroscopic redshifts within 10'' (shown in Figure~\ref{fig:split}) to estimate the redshifts of objects selected or rejected by the cut.

We select two intersecting lines in this color-color space, varying them systematically to optimize the resulting cut for our targeted redshift range.  First, we fit an initial guess line (Line 1) whose purpose is to minimize stellar contamination on the right-hand side of the plot.  Figure~\ref{split_by_stargal} shows the distribution of stars and galaxies in the color-color space, using the COSMOS star-galaxy classification described in~\S\ref{subsubsec:Zurich}.

Next, we fit a second line (Line 2) with some slope $m$, which will intersect with Line 1 at some value of $r-i$.  For each value of $m$, we can find the optimal $r-i$ of intersection between the Line 1 and Line 2, by varying the point of intersection from [-0.5, 2.0] and generating a ROC curve.  The optimal intersection between the two lines corresponds to the value of $r-i$ for which the distance from (0,1) in the ROC curve space is minimized.  We loop over the range of $m$, finding the best $r-i$ for each.  Slope $m$ varies clockwise, from just below the horizontal to vertical.  The optimal piecewise function for each slope tested, and their ROC curves, are shown in Figure~\ref{fig:ROC}.

To select the best classifier, we can compare the area under each ROC curve, as a better classifier will have an area closer to one.  The optimal color-color cut is the one whose area under the ROC curve is largest.  Lastly, we repeat this entire optimization process with Line 1 shifted slightly upward and downward, again comparing the area under the ROC curve to identify the best classifier.  The optimized classifier, targeting $z\ge0.55$, corresponds to the cut shown in Figure~\ref{split_by_stargal}: Line 2 with slope $m~=~0.249$, Line 1 shifted downward by 0.2 from our initial guess, and the point of intersection between Line 1 and Line 2 at $r-i~=~1.17$.

\begin{figure}
\begin{center}
\includegraphics[width=\columnwidth]{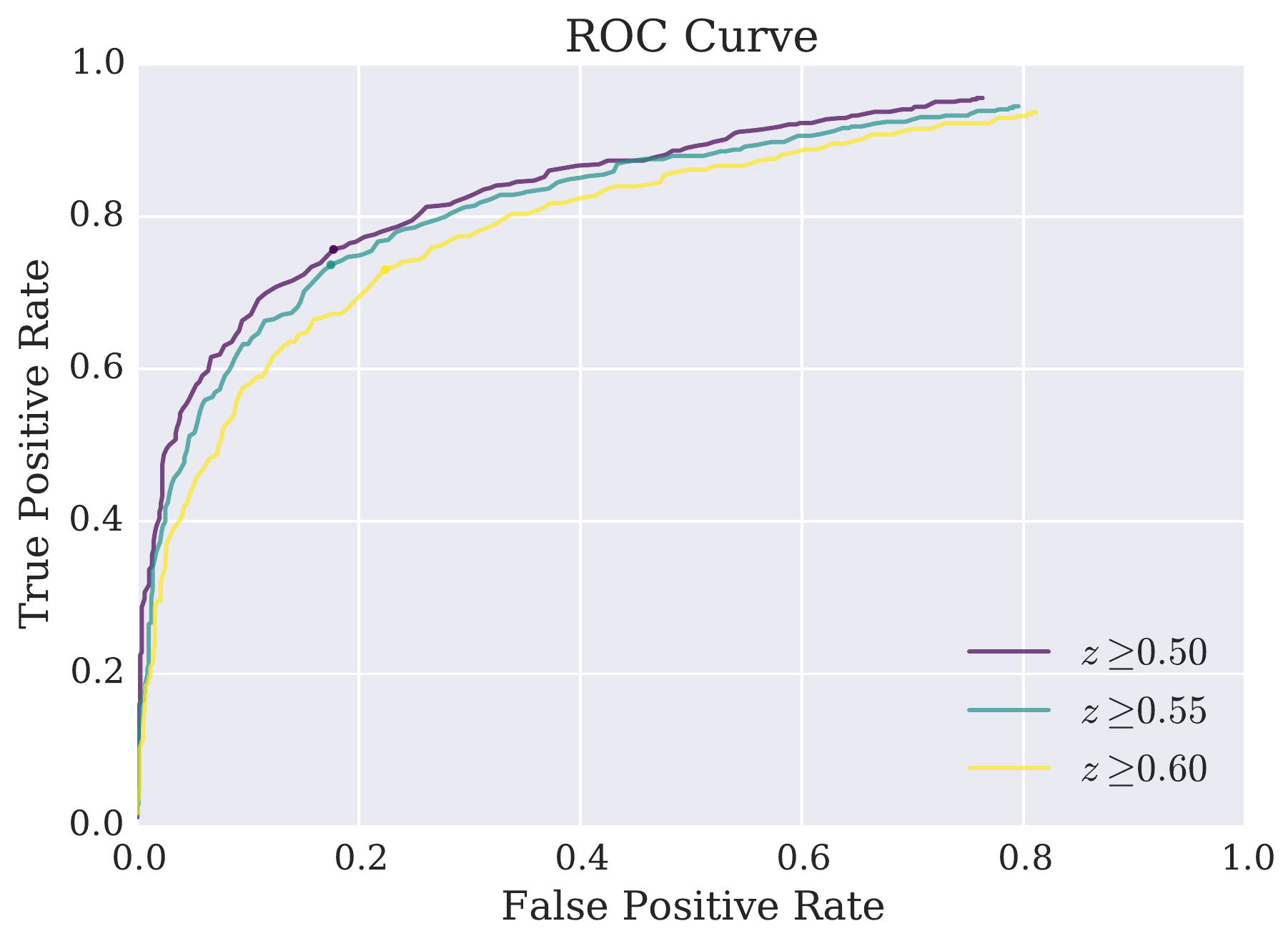}
\caption{ROC curve, showing the performance of the optimized selection color-color cut at various redshift requirements.  The cut performs slightly better for $z\ge0.50$.}
\label{thresholds}
\end{center}
\end{figure}

\begin{figure}
\begin{center}
\includegraphics[width=\columnwidth]{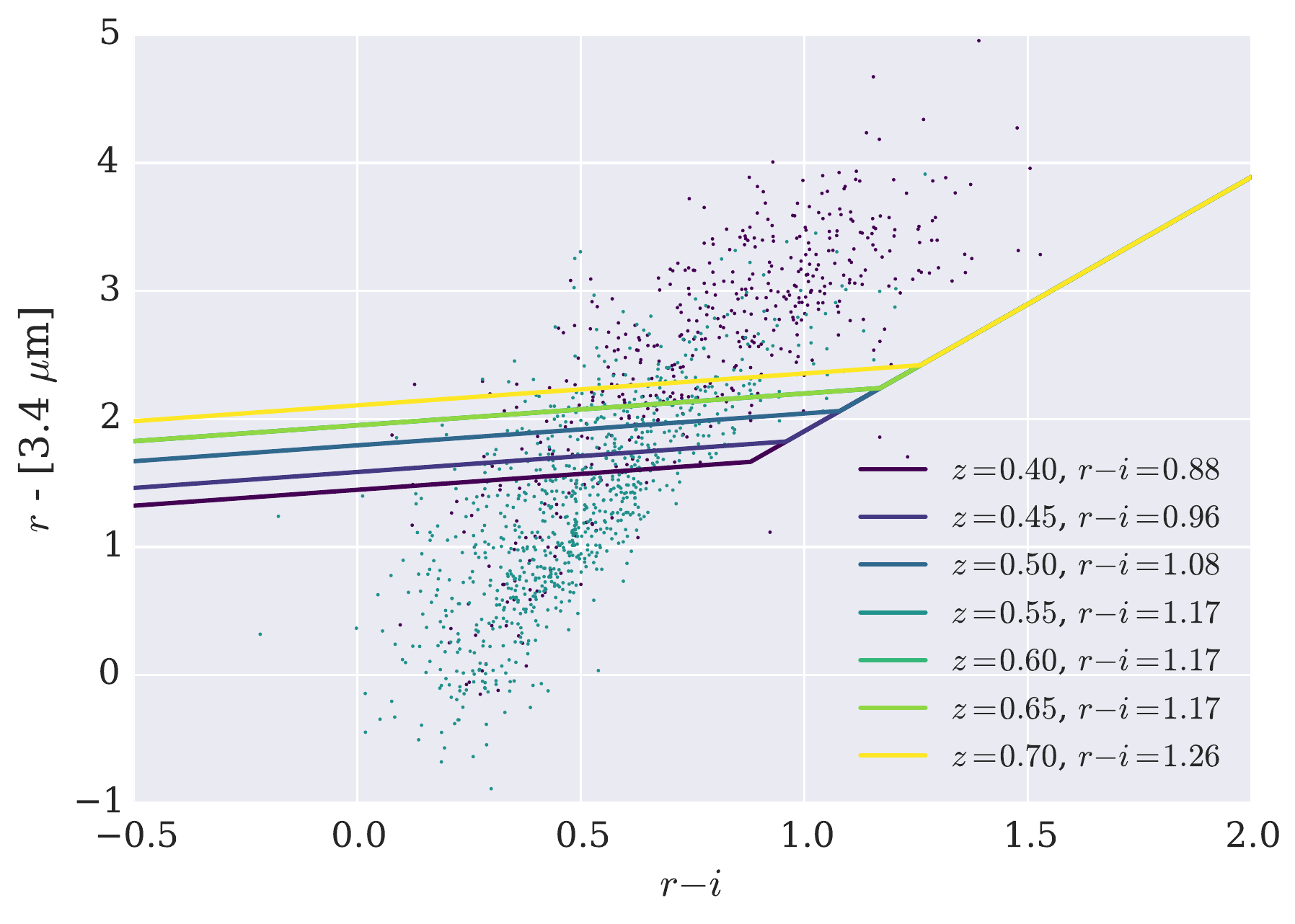}
\caption{Optimized color-color cut for different target redshifts.
The slope of Line 1 is held fixed at $m~=~0.249$, and the point of intersection between Line 1 and Line 2 is varied.
We show the targeted redshift and optimal point of intersection $r-i$ for each classifier.
Purple points are objects with $z\ge0.55$.
Teal points are objects with $z<0.55$.}
\label{target_z}
\end{center}
\end{figure}

We show the performance of this classifier against various redshift thresholds in Figure~\ref{thresholds}, and note that it performs slightly better for $z\ge0.5$.  We also show how the classifier varies if requiring different target redshifts in Figure~\ref{target_z}.  Lastly, we note that the area under the ROC curve is higher, i.e. the classifier performs better, if we also require targeted objects to have restframe $U-B>1.0$.  However, for the purposes of this work, we deliberately include bluer galaxies in addition to LRGs.

\begin{figure*}\begin{center}
\includegraphics[width=\textwidth]{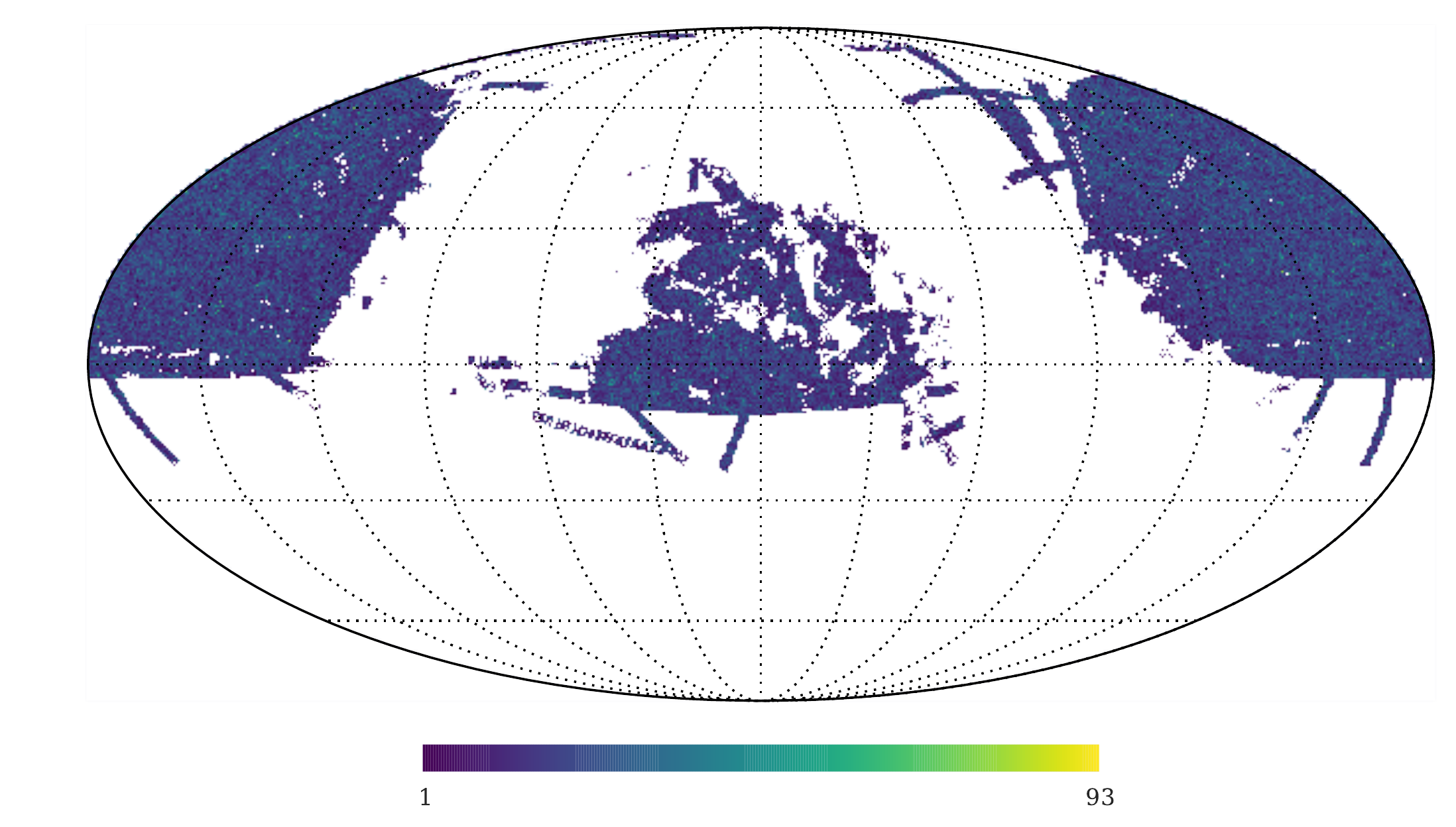}
\caption{Distribution of selected \numLRG LRG candidates along the full SDSS DR10 footprint.  Figure produced using HEALPix\footnote{\href{http://healpix.sf.net}{http://healpix.sf.net}} projection with $\textrm{NSIDE} = \nside$ (\resolution resolution), in equatorial coordinates.  Area per pixel is \pixArea.  Average density is \density selected objects per degree squared.}
\label{full_sky}
\end{center}\end{figure*}

\section{Properties of Selected Objects}

\label{sec:Results}

\numLRG massive galaxy candidates are selected over the full SDSS DR10 footprint, yielding a density of approximately \density objects per square degree.  Figure~\ref{full_sky} shows the distribution of these selected objects along the sky.

\subsection{Redshift Distribution}

\begin{figure*}
\begin{center}
\includegraphics[width=\textwidth]{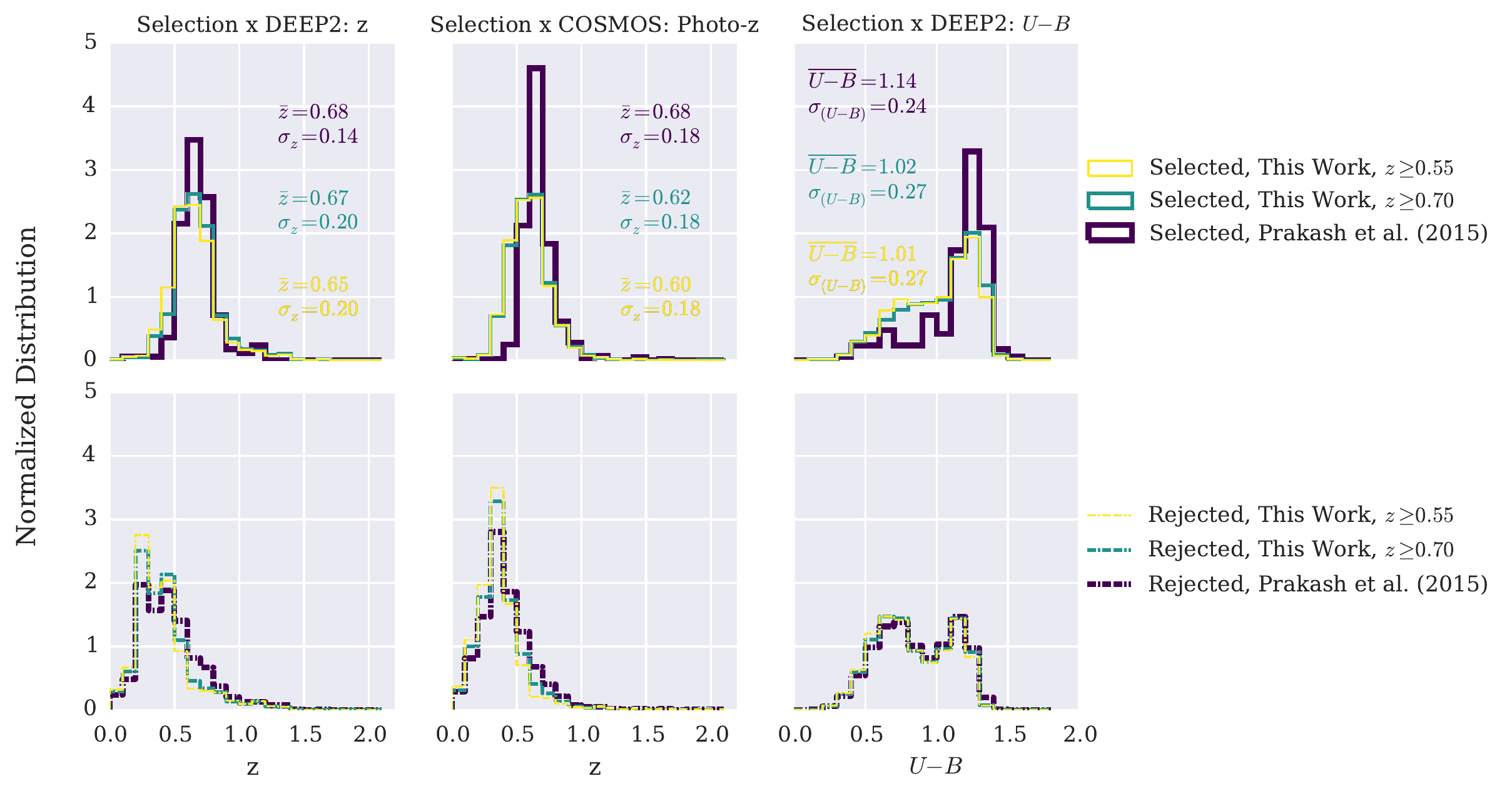}
\caption{Normalized redshift and restframe $U-B$ distributions of selected and rejected objects, and comparison to~\citet{Prakash:2015aa}.
The selection cut developed in this work provides a higher number of objects than the cut by~\citet{Prakash:2015aa}, and does not exclude bluer galaxies which are above the targeted redshift.
Left column:
Spectroscopic redshift distribution of objects cross-matched with the DEEP2 EGS field.  The histogram shows 531 selected objects targeting $z\ge0.55$, 438 selected objects targeting $z\ge0.70$, and 167 selected objects utilizing the cut proposed by~\citet{Prakash:2015aa}.
Center column:
Photometric redshift distribution of objects cross-matched with the COSMOS Photo-Z catalog.  The histogram shows 1793 selected objects targeting $z\ge0.55$, 1441 selected objects targeting $z\ge0.70$, and 407 selected objects utilizing the cut proposed by~\citet{Prakash:2015aa}.
Right column:
Restframe $U-B$ distribution of objects cross-matched with the DEEP2 EGS field.  The histogram shows 531 selected objects targeting $z\ge0.55$, 438 selected objects targeting $z\ge0.70$, and 167 selected objects utilizing the cut proposed by~\citet{Prakash:2015aa}.
Restframe $U-B$ distribution is used as an indicator of LRG selection efficiency.}
\label{Z_hist}
\end{center}
\end{figure*}

We cross-match the resulting catalog with both DEEP2 spectroscopic redshifts and COSMOS photometric redshifts within 10'' to test the efficiency of the cut.  The resulting normalized redshift distributions are shown in Figure~\ref{Z_hist}, alongside the redshift distribution of objects rejected by the selection method.

In the DEEP2 EGS field, we require cross-matched objects to lie within the 2D selection function map at values above $0.60$.  This allows for $82\%$ completeness for 434 cross-matching massive galaxy candidates.  The mean redshift is found to be $z=0.65$, with $\sigma=0.20$, and the median redshift is found to be $z=0.64$.   Contamination of selection rule for $z < 0.5$, $z<0.55$, and $z<0.6$ are $17\%$, $31\%$, and $42\%$, respectively.   Completeness of selection rule for $z>0.5$, $z>0.55$, and $z>0.6$ are $68\%$, $72\%$, and $72\%$, respectively.

In the COSMOS field, we are able to cross-match 1793 photometric redshifts to the selected galaxies, with $94\%$ completeness.  The mean redshift is found to be $z=0.60$, with $\sigma=0.18$, and the median redshift is found to be $z=0.60$.   Contamination of selection rule for $z < 0.5$, $z<0.55$, and $z<0.6$ are $27\%$, $42\%$, and $51\%$, respectively.   Completeness of selection rule for $z>0.5$, $z>0.55$, and $z>0.6$ are $70\%$, $72\%$, and $74\%$, respectively.

\subsection{$U-B$ Restframe Color}
\label{subsec:UB}

\begin{figure*}
\centering
\begin{tabular}{lll}
\centering\includegraphics[width=0.3\textwidth]{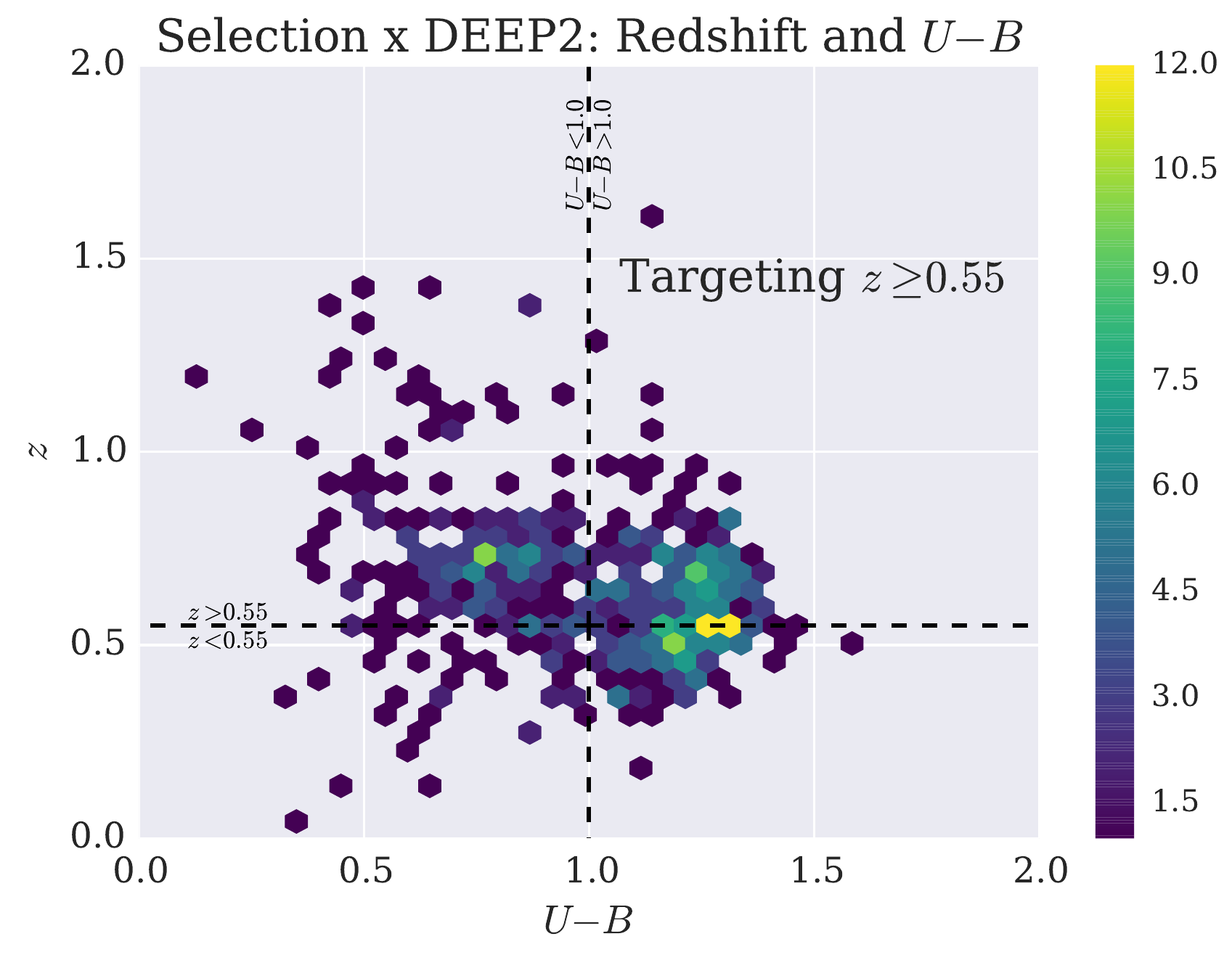} & \centering\includegraphics[width=0.3\textwidth]{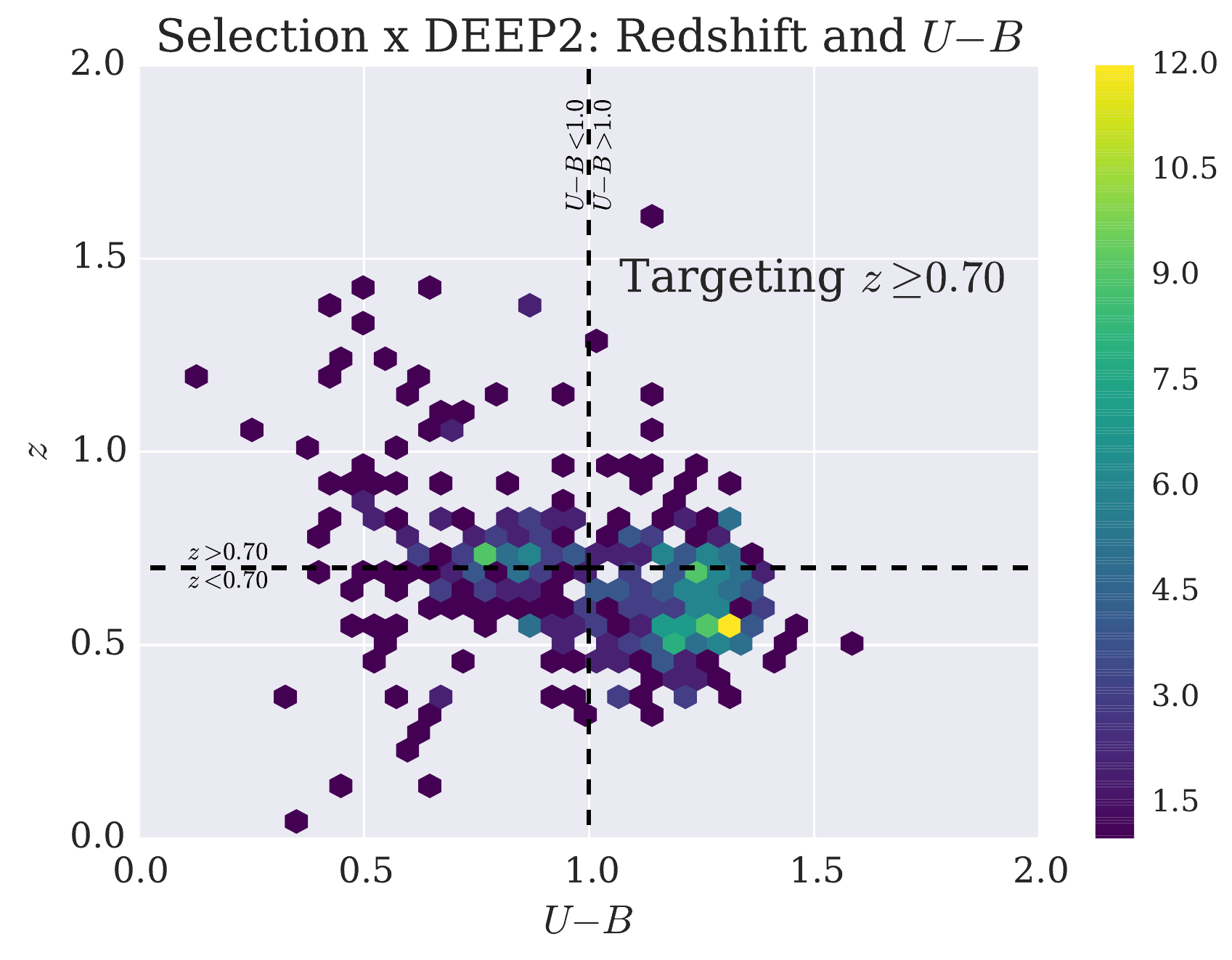} & \centering\includegraphics[width=0.3\textwidth]{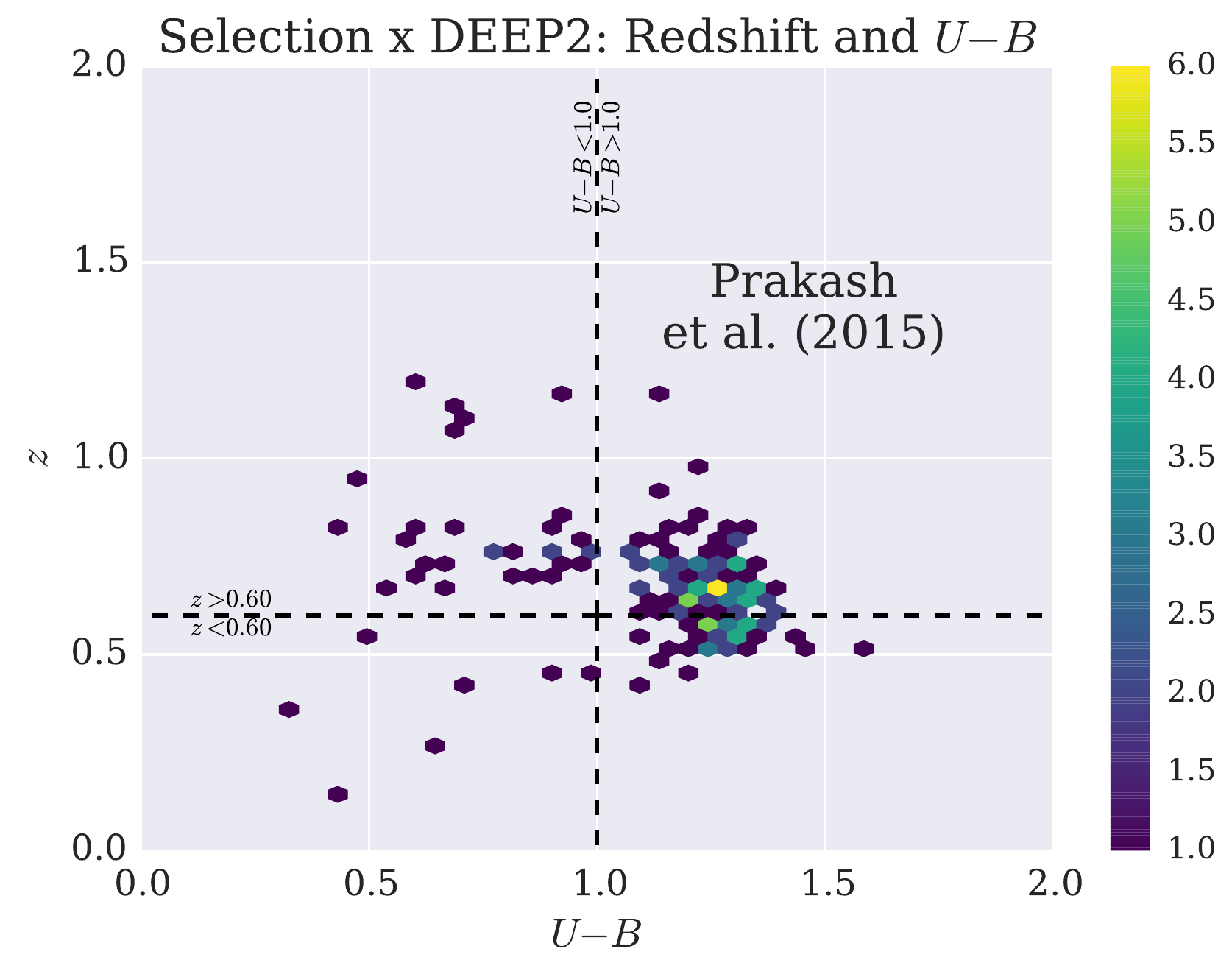} 
\end{tabular}
\caption{Two-dimensional histogram of redshift and restframe $U-B$ features of objects cross-matched with the DEEP2 EGS field.
The histograms show 531 selected objects targeting $z\ge0.55$ (left), 438 selected objects targeting $z\ge0.70$ (center), and 167 selected objects utilizing the cut proposed by~\citet{Prakash:2015aa} (right).
In the selection cut targeting galaxies with $z\ge 0.55$, only $9.6\%$ of the contamination comes from bluer sources with $U-B<1.0$ and redshift $z<0.55$}
\label{Z_UB}
\end{figure*}

We use restframe $U-B$ colors contained in the DEEP2 Redshift Galaxy Survey catalog to determine LRG selection efficiency.  Figure~\ref{Z_hist} shows the distribution of $U-B$ for selected and rejected objects. Average restframe $U-B=1.0$, with $\sigma=0.27$, and median $U-B=1.1$.  Blue galaxies make up an estimated $44\%$ of the catalog.  Figure~\ref{Z_UB} shows the targeted properties of redshift and $U-B$ plotted together.  Blue sources below the target redshift make up an estimated $9.6\%$ of the catalog.

\subsection{Stellar Contamination}
Stellar contamination is estimated using the $ACS\_MU\_CLASS$ flag contained within the COSMOS Zurich Structure \& Morphology Catalog described in~\S\ref{subsubsec:Zurich}.  Over the COSMOS area, 1,648 objects are cross-matched to within 10'' of the selected objects, with a completeness of $98\%$.  We estimate a stellar contamination of $1.8\%$.

\subsection{Comparison to Previous Work}
\label{subsec:Prak}

We include, for comparison, the selection cut developed by~\citet{Prakash:2015aa}, shown in Figure~\ref{split_by_stargal}.  The cut performs best for a threshold $z\ge0.6$, and yields a more peaked redshift distribution closer to $z=0.68$.  In Figure~\ref{Z_hist} we show the resulting redshift distribution by~\citet{Prakash:2015aa} applied to our data.  However, as seen in Figure~\ref{Z_hist}, the cut presented in this work does not reject as many massive blue galaxies.  Because the cut is wider, we are also able to select a much larger number of galaxies.

\section{Conclusion and Future Work}
\label{sec:Conclusion}

We have efficiently selected a catalog of massive galaxies optimized to select objects with $z\ge0.55$, using optical and infrared photometry.  In DEEP2, the resulting catalog has average redshift $z=0.65$, with standard deviation $\sigma = 2.0$.  In COSMOS, the resulting catalog has average redshift $z=0.60$, with standard deviation $\sigma = 1.8$.  Average restframe $U-B=1.0$, with $\sigma=0.27$.  The catalog contains primarily LRGs, although an estimated $44\%$ of the selected objects are bluer galaxies.  We anticipate these to be massive galaxies in our targeted redshift range that will be equally cosmologically useful.  We find that only $9.6\%$ of the catalog are bluer sources with redshift $z<0.55$.  Moreover, the selection yields a higher number of galaxies than previous work by~\citet{Prakash:2015aa}.  Stellar contamination is estimated to be $\stellarContamination\%$.
 
We anticipate a large signal from cross-correlations with CMB lensing, detections of the Sunyaev-Zel'dovich effect, and the Integrated Sachs-Wolfe effect.  The catalog will be publicly available\footnote{\href{https://sites.google.com/site/massivegalaxycatalog}{https://sites.google.com/site/MassiveGalaxyCatalog}}.

\section{Acknowledgements}
\label{sec:Acknowledgements}


This work was begun as an undergraduate senior thesis at Princeton University, under the supervision of David N. Spergel and with the support of NSF grant AST-1311756.  This work was partially funded by a Fulbright grant under the U.S. Student Program to Chile.  We would like to thank Rolando D{\"u}nner for his many useful discussions throughout the development of this project.  This work was also supported by NASA 12-EUCLID11-0004 and NSF AST1517593, under the supervision of Shirley Ho.  We would like to thank David Wake for his contributions in the early stages of this work.  Lastly, we would like to thank Dustin Lang for his help in accessing the unWISE data.


Funding for the DEEP2 Galaxy Redshift Survey has been provided by NSF grants AST-95-09298, AST-0071048, AST-0507428, and AST-0507483 as well as NASA LTSA grant NNG04GC89G.


Some of the results in this paper have been derived using the HEALPix~\citep{Gorski:2005aa} package.

Figures in this paper have been developed using formatting from the Seaborn~\citep{Seaborn} package.


Funding for the SDSS and SDSS-II has been provided by the Alfred P. Sloan Foundation, the Participating Institutions, the National Science Foundation, the U.S. Department of Energy, the National Aeronautics and Space Administration, the Japanese Monbukagakusho, the Max Planck Society, and the Higher Education Funding Council for England. The SDSS Web Site is http://www.sdss.org/.

The SDSS is managed by the Astrophysical Research Consortium for the Participating Institutions. The Participating Institutions are the American Museum of Natural History, Astrophysical Institute Potsdam, University of Basel, University of Cambridge, Case Western Reserve University, University of Chicago, Drexel University, Fermilab, the Institute for Advanced Study, the Japan Participation Group, Johns Hopkins University, the Joint Institute for Nuclear Astrophysics, the Kavli Institute for Particle Astrophysics and Cosmology, the Korean Scientist Group, the Chinese Academy of Sciences (LAMOST), Los Alamos National Laboratory, the Max-Planck-Institute for Astronomy (MPIA), the Max-Planck-Institute for Astrophysics (MPA), New Mexico State University, Ohio State University, University of Pittsburgh, University of Portsmouth, Princeton University, the United States Naval Observatory, and the University of Washington.

Funding for SDSS-III has been provided by the Alfred P. Sloan Foundation, the Participating Institutions, the National Science Foundation, and the U.S. Department of Energy Office of Science. The SDSS-III web site is http://www.sdss3.org/.

SDSS-III is managed by the Astrophysical Research Consortium for the Participating Institutions of the SDSS-III Collaboration including the University of Arizona, the Brazilian Participation Group, Brookhaven National Laboratory, Carnegie Mellon University, University of Florida, the French Participation Group, the German Participation Group, Harvard University, the Instituto de Astrofisica de Canarias, the Michigan State/Notre Dame/JINA Participation Group, Johns Hopkins University, Lawrence Berkeley National Laboratory, Max Planck Institute for Astrophysics, Max Planck Institute for Extraterrestrial Physics, New Mexico State University, New York University, Ohio State University, Pennsylvania State University, University of Portsmouth, Princeton University, the Spanish Participation Group, University of Tokyo, University of Utah, Vanderbilt University, University of Virginia, University of Washington, and Yale University.


This publication makes use of data products from the Wide-field Infrared Survey Explorer, which is a joint project of the University of California, Los Angeles, and the Jet Propulsion Laboratory/California Institute of Technology, and NEOWISE, which is a project of the Jet Propulsion Laboratory/California Institute of Technology. WISE and NEOWISE are funded by the National Aeronautics and Space Administration.

\clearpage

\end{document}